\documentclass[letterpaper,10pt,nofootinbib,aps,tightenlines,twocolumn]{revtex4}

\usepackage{amsmath,amsfonts,amssymb}
\usepackage{mathrsfs}
\usepackage{graphicx}
\usepackage[english]{babel} 

\usepackage{appendix}

\usepackage{epstopdf}

\begin{document}

%%%%%%%%%%%%%%%%%%%%%%%%%%%%%%%%%%%%%%%%%%%%%%%%%%%%%%%%%%%
\title{A Brief History of Curvature}

\author{Robert R. Caldwell$^1$ and Steven S. Gubser$^2$}
\affiliation{$^1$Department of Physics \& Astronomy, Dartmouth College, 6127 Wilder Laboratory, Hanover, NH 03755 USA}
\affiliation{$^2$Joseph Henry Laboratories, Princeton University, Princeton, NJ 08544, USA}

\date{\today}

%%%%%%%%%%%%%%%%%%%%%%%%%%%%%%%%%%%%%%%%%%%%%%%%%%%%%%%%%%%
\begin{abstract}
\begin{picture}(0,0)(0,0)\put(397,120){PUPT-2439}\end{picture}
The trace of the stress-energy tensor of the cosmological fluid, proportional to the Ricci scalar curvature in general relativity, is determined on cosmic scales for times ranging from the inflationary epoch to the present day in the expanding Universe. The post-inflationary epoch and the thermal history of the relativistic fluid, in particular the QCD transition from asymptotic freedom to confinement and the electroweak phase transition, leave significant imprints on the scalar curvature.  These imprints can be of either sign and are orders of magnitude larger than the values that would be obtained by naively extrapolating the pressureless matter of the present epoch back into the radiation-dominated epoch. 
\end{abstract}
%%%%%%%%%%%%%%%%%%%%%%%%%%%%%%%%%%%%%%%%%%%%%%%%%%%%%%%%%%%
\maketitle

%%%%%%%%%%%%%%%%%%%%%%%%%%%%%%%%%%%%%%%%%%%%%%%%%%%%%%%%%%%
\section{Introduction}

The broad history of the Universe, according to the Standard Cosmological Model, is a succession of expansion epochs from the earliest moments of the Big Bang to the present day. In reverse chronological order, these are: dark-energy domination, during which something like a cosmological constant drives the accelerated cosmic expansion; matter-domination, during which cold dark matter and baryons shape the large-scale structure of the Universe; radiation-domination, during which the properties of the cosmic fluid are described by the relativistic degrees of freedom of the Standard Model (SM) of particle physics and beyond; and inflation, during which the potential energy of a scalar degree of freedom dominates the cosmic energy density, producing an exponential burst of cosmic expansion. Despite fundamental uncertainties---the nature of dark energy and dark matter, and the physics beyond the Standard Model remain open questions, whilst the search continues for direct evidence of the inflationary birth of the Universe---this sequence of epochs forms the architecture of a widely-accepted, concordant model of the Universe.

The aim of this article is to give a more quantitatively accurate account of cosmic history than one obtains from patching together inflationary, radiation-, matter-, and dark energy-dominated epochs. It may often suffice to describe the cosmic energy density $\rho \propto a^{-3(1+w)}$ in terms of the scale factor $a$ and the equation of state $w =(-1,\, 1/3,\, 0,\, -1)$ for inflation, radiation, matter, and dark energy. However, it is rare to find a detailed history of the trace of the stress-energy tensor of the cosmological fluid, or equivalently the Ricci scalar curvature, $R = \kappa \Theta$ where $\kappa \equiv 8 \pi G$, and $\Theta \equiv -T^\mu_\mu = \rho-3p$.\footnote{Conventions: the metric has signature $-++\,+$, the Einstein Field Equations are $G^{\mu\nu}=\kappa T^{\mu\nu}$, with $R_{\mu\nu} = \Gamma^\lambda_{\mu\nu,\lambda}+...$ and $T^{\mu\nu}=(\rho+p)u^\mu u^\nu + p g^{\mu\nu}$. The variable $\Theta$ is introduced to avoid confusing the trace with temperature $T$.} The final output of our calculations is a determination of $R$ and $\Theta$, as functions of cosmic time or redshift: See Fig.~\ref{fig:TraceFullHist}.

To make these calculations, we must properly account for the thermal history of the cosmological fluid, the dynamics of the phase transition of quantum chromodynamics (QCD), and we must give some account of the electroweak phase transition. We do not solve the full Boltzmann equations, as necessary for the standard calculations of the light element abundances in Big Bang Nucleosynthesis and the relic dark matter abundance in extensions of the Standard Model (e.g. \cite{Kolb:1990vq}), but instead use free field thermodynamics where applicable. We also make some assumptions about inflation and reheating in the early Universe in order to follow $R$ and $\Theta$ all the way back to the inflationary epoch. 

It is often supposed that the trace of the stress-energy tensor is negligibly small during the radiation-dominated epoch.  It is also supposed that the dominant contribution to the trace comes from baryons and dark matter.  Both of these suppositions are incorrect: massive particle species contribute non-negligibly to the trace when the temperature is comparable to their mass (see e.g. Refs.~\cite{Damour:1994zq,Cembranos:2009ds}, wherein the history of $\Theta$ is required to study cosmic evolution in a scalar-tensor theory of gravitation), and the effect of phase transitions on the trace of the stress-energy tensor is in fact much larger than the small contribution from baryons and dark matter.

Our motivation is to chart a detailed history of $\Theta$ as comprehensively as we can.  Two advances outside of cosmology make this possible.  First, progress in lattice QCD shows that the self-interactions of the quark-gluon plasma produce a trace that is very different than what one would obtain using free field theory \cite{Bazavov:2009zn,Borsanyi:2010cj}.  Second, the Higgs mass is now known---or at least, the mass of a particle whose most straightforward interpretation is the Standard Model Higgs \cite{ATLAS:2012gk,CMS:2012gu}.  Because we don't know what lies beyond the Standard Model, we offer two scenarios.  In the first scenario (despite all arguments about naturalness) nothing lies beyond the Standard Model until one reaches the energy scale of inflation, only somewhat below the typical scale of Grand Unified Theories.  In the second scenario, there is a supersymmetric extension of the Standard Model plausibly within the reach of discovery at the LHC, and then inflationary physics at a similar energy scale to before.  Our final figure covers the entire history of the Universe from inflation to the present day for both scenarios.

\begin{figure*}[t]
\begin{center}
\includegraphics[scale=0.6]{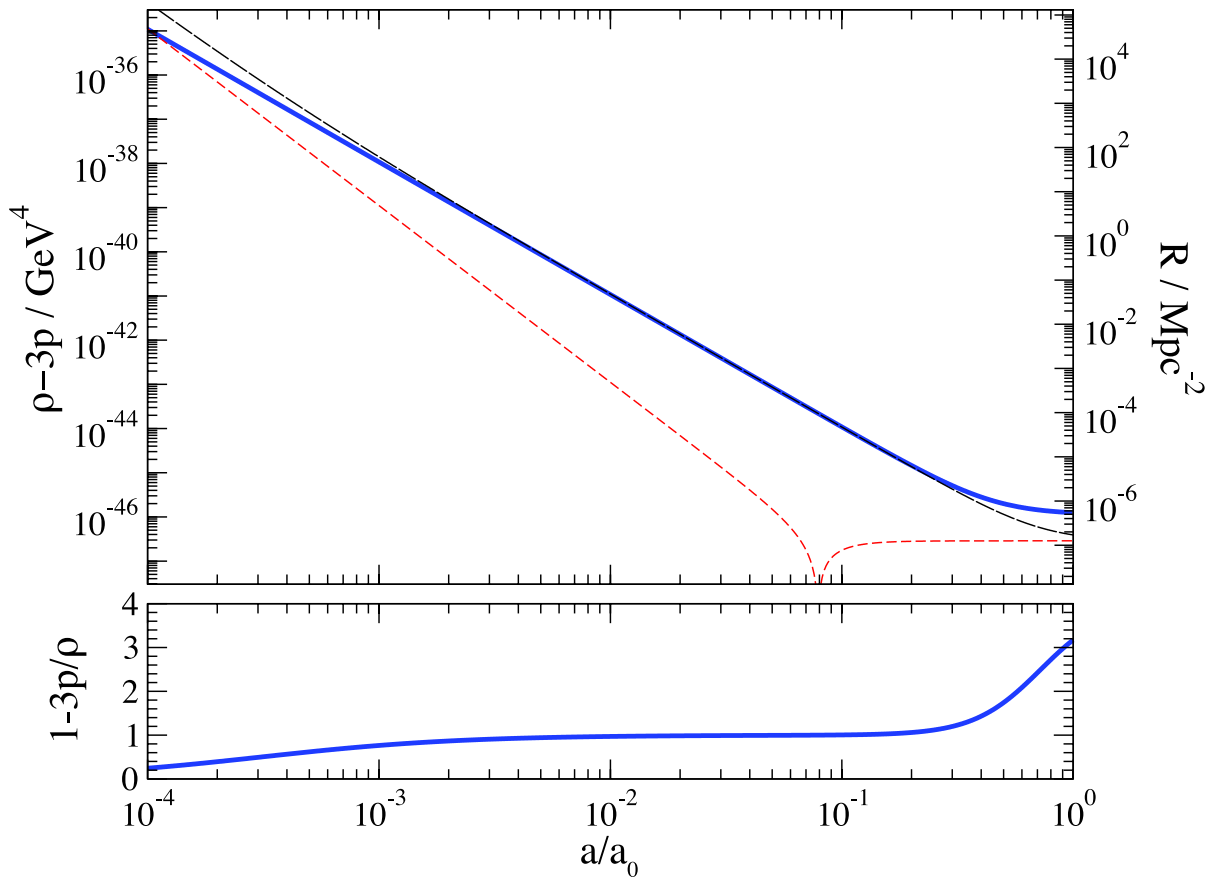}
\end{center}
\caption{The Ricci scalar curvature in units of ${\rm Mpc}^{-2}$, or equivalently the trace of the cosmological fluid stress-energy tensor in units of ${\rm GeV}^4$, is shown by the thick (blue) curve for $10^{-4} < a/a_0 < 1$ or a run of redshift $0 <z <10^4$. At late times, the trace approaches a constant value due to the onset of dark energy domination. For reference, the energy density is shown by the thin, long dashed (black) curve. The pressure is shown by the thin, short dashed (red) curve. Note that the pressure is negative after $a/a_0 \gtrsim 0.08$, so the absolute value is plotted. In the lower panel, the trace in units of the total energy density is shown, $1 - 3 p/\rho$.}
\label{fig:TraceMattEra}
\end{figure*}

%%%%%%%%%%%%%%%%%%%%%%%%%%%%%%%%%%%%%%%%%%%%%%%%%%%%%%%%%%%
\section{The matter-dominated epoch and afterwards}

A good (and, by now, entirely conventional) account of the expanding universe for $z \lesssim 100$ can be achieved by including only non-relativistic baryons, pressureless cold dark matter, and dark energy with $w=-1$.  Of course photons and neutrinos are present, but their contributions to the stress-energy tensor are negligible at this stage. As yet, there is no significant evidence for time-evolving dark energy (e.g. Ref.~\cite{Zhao:2012aw}); however, the inferred, model-independent evolution at $z \lesssim 1$ remains the subject of intense investigation as a measure of the dark energy equation-of-state.  Thus $w=-1$ for dark energy is only a default assumption.

The trace of the stress-energy tensor, and therefore the Ricci scalar curvature, during this epoch is therefore
\begin{eqnarray}
\label{eqn:mattercurv}
R = \kappa \Theta &=& \kappa( \rho_{c} + \rho_{b} + (1 - 3w_{de})\rho_{de} ) \\
&=& 3 H_0^2 \left(( \Omega_{c}+\Omega_{b}) (a_0/a)^3 +4 \Omega_{de}\right), \nonumber
\end{eqnarray}
as illustrated in Fig.~\ref{fig:TraceMattEra}. The central values of the WMAP-derived cosmological parameters---$\Omega_b h^2 = 0.02249$ for baryons, $\Omega_{c} h^2 =0.1120$ for cold dark matter, and $\Omega_{de} = 0.727$ \cite{Larson:2010gs} for dark energy---are used to set  $\rho_{de} = \Omega_{de} \rho_{0}$, $\rho_{c} = \Omega_{c} \rho_{0} (a_0/a)^3$, $\rho_{b} = \Omega_{b} \rho_{0} (a_0/a)^3$ and $\rho_0 = 3 H_0^2/8 \pi G$ where $a$ is the expansion scale factor.  The Hubble constant is $H_0 = 100 h$~km/s/Mpc with $h = 0.738 \pm  0.024$ \cite{Riess:2011yx}. For the moment we neglect any non-relativistic neutrino species, whose contribution is small relative to the baryons and dark matter. We also ignore all fluctuations and effects of structure formation: our purpose is only to find the Robertson-Walker (RW) background that best approximates the expanding Universe at very large scales.

%%%%%%%%%%%%%%%%%%%%%%%%%%%%%%%%%%%%%%%%%%%%%%%%%%%%%%%%%%%
\section{Free field treatment of the radiation-dominated epoch}

The free field treatment of the radiation-dominated epoch is based on the following standard formulas for a collection of non-interacting fermions and bosons in thermal equilibrium:
 \begin{eqnarray}
  \rho_{\rm free} &=& \sum_j {g_j \over 2\pi^2} \int_{m_j}^\infty dE \,
    {E^2 \sqrt{E^2 - m_j^2} \over e^{E/T_j} - s_j}  \label{FreeE} \\
  p_{\rm free} &=& \sum_j {g_j \over 6\pi^2} \int_{m_j}^\infty dE \,
    {(E^2-m_j^2)^{3/2} \over e^{E/T_j} - s_j} \,,  \label{FreeP}
 \end{eqnarray}
where the sum is over all species of particle,  $m_j$ is the mass of the $j$-th species, $g_j$ is its multiplicity, and $s_j = +1$ for bosons and $-1$ for fermions \cite{Kolb:1990vq}.  We have allowed for different species to have different temperatures $T_j$.  For each collection of mutually thermally equilibrated particles, the common temperature must evolve in such a way that the collection's contribution to the stress-energy tensor is conserved.

It is not necessary to generalize (\ref{FreeE}) and (\ref{FreeP}) to include chemical potentials, because all the relevant chemical potentials (for example, for baryon number) are small compared to the temperature at times when explicit expressions for non-zero pressure are needed.  

\begin{figure}[h]
\begin{center}
\includegraphics[scale=0.33]{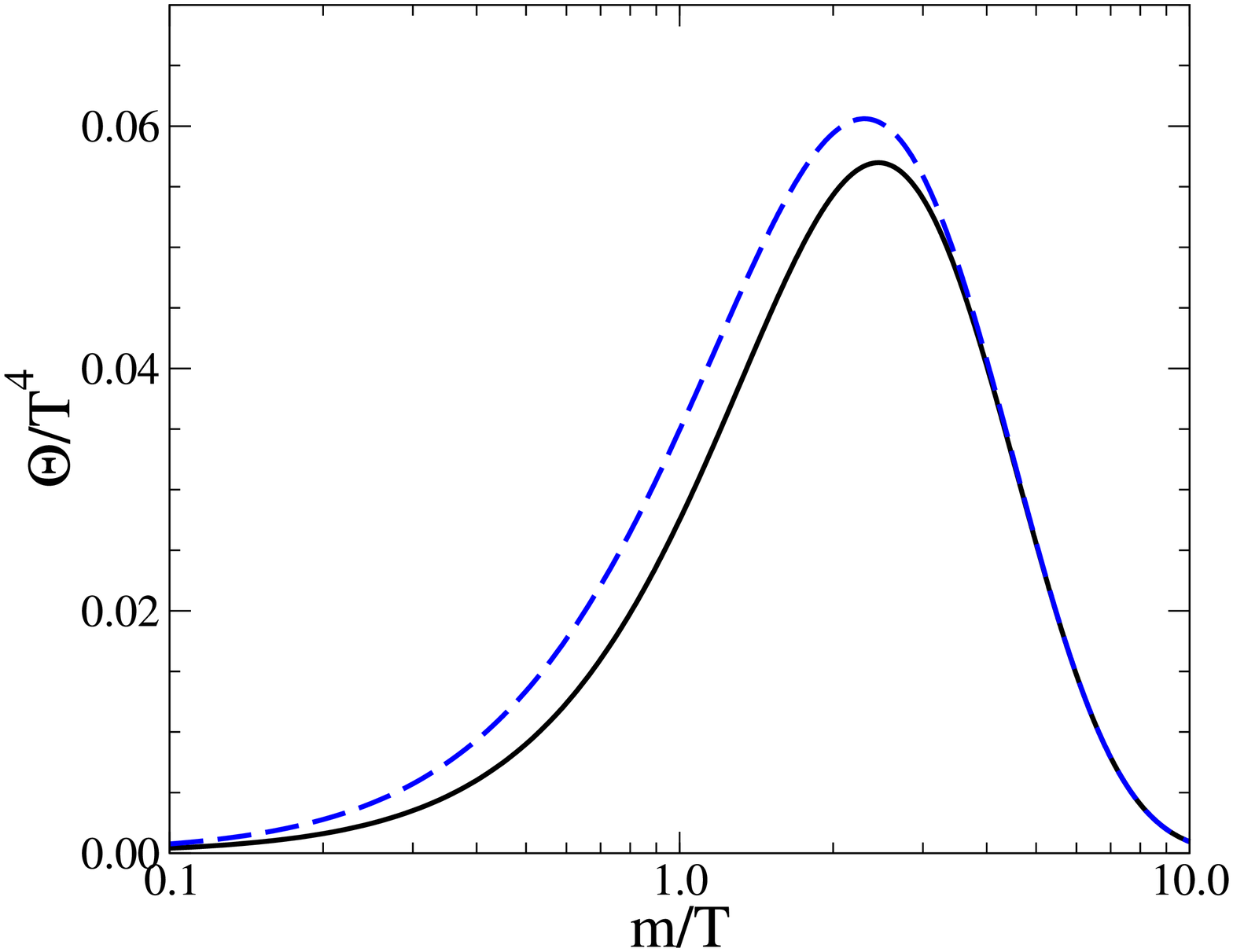}
\end{center}
\caption{The contribution to the ratio $\Theta/T^4$ by a species ($g=1$) cooling below its rest mass is shown. The solid (dashed) line is for fermions (bosons). According to a free field treatment, the freeze-out of each species of the Standard Model results in a bump to the trace.}
\label{fig:CurvatureSpike}
\end{figure}

The contribution to the trace $\Theta$ by photons is identically zero, whereas a non-zero trace is obtained for any free, massive species. In the relativistic limit $m \ll T$ the leading contribution to the trace is 
\begin{equation}
\Theta \simeq 
\begin{cases} \frac{g}{24} m^2 T^2  & {\rm fermion,} \\
\frac{g}{12} m^2 T^2  & {\rm boson.}
\end{cases}
\end{equation}
This means that as long as the relativistic fluid includes a massive species, then the trace, and therefore the curvature, is non-vanishing. The energy density and pressure decay at slightly different rates as a species cools, so that as the temperature drops past about half the mass, a bump occurs in the contribution to the trace. As an illustrative example, Fig.~\ref{fig:CurvatureSpike} shows the ratio $\Theta/T^4$ for a single boson or fermion species cooling below its rest mass. We expect this process to be repeated for each relativistic species of the Standard Model, briefly raising the trace above a background level set by any abundant, non-relativistic particles for which $\Theta = \rho$, or else the most massive, relativistic species in equilibrium for which $\Theta \propto m^2 T^2$. 

The most recent example of this phenomena is due to electrons as the temperature of the cosmic fluid drops below the electron mass. Turn ahead to Fig.~\ref{fig:TraceRadEra} to see this illustrated. Whilst Standard Model neutrinos become non-relativistic more recently, their contribution to the trace is negligible.\footnote{The sum of the neutrino masses is constrained to lie in the range $0.06~{\rm eV} < \sum m_\nu < 1.1$~eV at $95\%$~CL, where the lower bound is due to the mass splittings seen in neutrino oscillation experiments \cite{Nakamura:2010zzi} and the upper bound is obtained from cosmological observations \cite{Komatsu:2010fb,Abazajian:2011dt}. For these masses, neutrinos become non-relativistic at a time when the radiation energy density is smaller than the contribution due to baryons and cold dark matter. Consequently, the peak ratio of the trace contributed by neutrinos to non-relativistic matter is found to be $\Theta_\nu / \Theta_m \approx 0.02\, (\sum m_\nu / {\rm eV})$ occuring at a redshift $z \approx 800\, (\sum m_\nu / {\rm eV})$. Hence, this is at most a $2\%$ change in the cosmic trace. This is comparable to other phenomena which shift the effective number of relativistic neutrino species upwards from $3$ to $3.046$ which we also neglect. See Ref.~\cite{Lesgourgues:2006nd} for a discussion of these corrections.} Consequently, all one needs to model the cosmic trace at temperatures $T \lesssim 10^{-4}$GeV or for $a/a_0 \gtrsim 10^{-9}$, long before the onset of the matter-dominated era, is baryons, cold dark matter, and dark energy as given by Eq.~(\ref{eqn:mattercurv}). At temperatures $T \gtrsim 10^{-4}$GeV or $a/a_0 \lesssim 10^{-9}$, a virtual parade of particle species, each cooling past their individual rest mass energies, keeps the cosmic trace well above the level of the energy density in cold dark matter.  
 
Proceeding to higher temperatures, it is convenient to describe the relativistic gas in terms of an effective number of degrees of freedom $g_*, \, g_{*p}$ for the energy density and pressure
\begin{equation}
  \rho = g_* \frac{\pi^2}{30} T^4,   \qquad
  p = g_{*p} \frac{\pi^2}{90} T^4.  \label{eqn:gstar}
\end{equation}
The stress-energy trace is $\Theta = (g_* - g_{*p}){\pi^2}T^4/ 30$. For the free, massive particles of the Standard Model composing the cosmological fluid it is straightforward to use equilibrium thermodynamics to calculate $g_*,\,g_{*p}$.  As we pass to larger temperatures and redshifts, however, the main questions are:  (1)  When can we use a free field treatment of thermodynamics?
(2) When free field thermodynamics fails, what should we use instead?
We will consider four cases in the following sections: the QCD phase transition, where lattice results are available and are significantly different from free field results; the Standard Model electroweak transition, where ring-improved one-loop calculations indicate that departures from free field expectations are not too great; a supersymmetric extension of the Standard Model, where we {\it assume} that the effect of interactions on the thermodynamics is fairly modest; and finally inflation and reheating.
 
%%%%%%%%%%%%%%%%%%%%%%%%%%%%%%%%%%%%%%%%%%%%%%%%%%%%%%%%%%%
\section{The QCD phase transition}
\label{QCD}

Mesons and baryons melt into constituent quarks and gluons at the QCD phase transition. The best source of information about thermodynamics in the vicinity of the transition is lattice QCD: see for example Refs.~\cite{Bazavov:2009zn,Borsanyi:2010cj}.  While the results of these works are qualitatively similar, quantitatively they differ enough so that it is best to summarize each separately.

The authors of Ref.~\cite{Bazavov:2009zn} provide the following parametrized form:
 \begin{equation}
  {\Theta \over T^4} = \left[ 1 - {1 \over (1+e^{(T-c_1)/c_2})^2} \right]
    \left( {d_2 \over T^2} + {d_4 \over T^4} \right) \,.  \label{BazavovForm}
 \end{equation}
We will use the values $c_1 = 193.8\,{\rm MeV}$, $c_2 = 13.6\,{\rm MeV}$, $d_2 = 0.241\,{\rm GeV}^2$, and $d_4 = 0.0035\,{\rm GeV}^4$ given in \cite{Bazavov:2009zn}, which include a $10\,{\rm MeV}$ shift downward in temperature of the lattice data, which slightly reduces the discrepancy with \cite{Borsanyi:2010cj} (to be described in the next paragraph).  The pressure as a function of $T$ can be extracted from the general result
 \begin{equation}
  {p(T) \over T^4} - {p(T_0) \over T_0^4} = \int_{T_0}^T dT' \, 
    {\Theta(T') \over T'^5} \,,  \label{pFromTheta}
 \end{equation}
together with some information about $p$ at low temperatures.  In order to find $\rho(T)$ and $p(T)$ at low temperatures, we used the free field expressions (\ref{FreeE}) and (\ref{FreeP}) with the sum running over pions, kaons, and the $\eta$, $\rho$, and $\omega$ mesons, together with protons, neutrons, and their anti-particles.  The free field value of $\Theta/T^4$ matches the one given in (\ref{BazavovForm}) at $T_0=134\,{\rm MeV}$.  (It is coincidental that this is almost exactly the mass of the neutral pion.)  For $T>T_0$ we extracted $p$ from (\ref{pFromTheta}) and then $\rho$ from (\ref{BazavovForm}).

The authors of \cite{Borsanyi:2010cj} provide the following parametrized form as a fit to their results:
 \begin{equation}
  {\Theta \over T^4} = e^{-{h_1 \over t} - {h_2 \over t^2}} \left[
     h_0 + f_0 {\tanh(f_1 t + f_2) \over 1 + g_1 t + g_2 t^2} \right] \,,  \label{FodorForm}
 \end{equation}
where $t=T/(200\,{\rm MeV})$ and $h_0 = 0.1396$, $h_1 = -0.1800$, $h_2 = 0.0350$, $f_0 = 2.76$, $f_1 = 6.79$, $g_1 = -0.47$, and $g_2 = 1.04$.  The relation (\ref{pFromTheta}) can again be used to extract $p(T)$.  The form (\ref{FodorForm}) describes the lattice data from $T=100\,{\rm MeV}$ to $1\,{\rm GeV}$, and it also is a good fit to the free field thermodynamics from $T=20\,{\rm MeV}$ to $T=100\,{\rm MeV}$.  Thus we can use (\ref{FodorForm}) essentially without additional low-temperature information---or match to free field thermodynamics at $T_0 = 20\,{\rm MeV}$. 

We note that a fairly good fit to the lattice data of \cite{Borsanyi:2010cj}, which covers the range $100\,{\rm MeV} < T < 1\,{\rm GeV}$, can be achieved with a simpler functional form:
 \begin{equation}
  {\Theta \over T^4} = \left[ 1 - {1 \over (1+e^{(T-c_1)/c_2})^2} \right] {d_{3/2} \over T^{3/2}} \,,
    \label{ThreeHalvesForm}
 \end{equation}
where $c_1 = 180\,{\rm MeV}$, $c_2 = 20\,{\rm MeV}$, and $d_{3/2} = 0.39\,{\rm GeV}^{3/2}$.  The $1/T^{3/2}$ scaling is curious because it suggests a leading dimension $\Delta \approx 5/2$ for the operator which breaks conformal invariance at higher temperatures.  The approximate $1/T^{3/2}$ scaling persists over less than a factor of $10$ on the temperature axis, so we should not take this scaling dimension too seriously.  No free field construction will give a scalar operator with $\Delta = 5/2$, but below $1\,{\rm GeV}$ we are far from asymptotic freedom.

\begin{figure}[t]
\begin{center}
\includegraphics[scale=0.33]{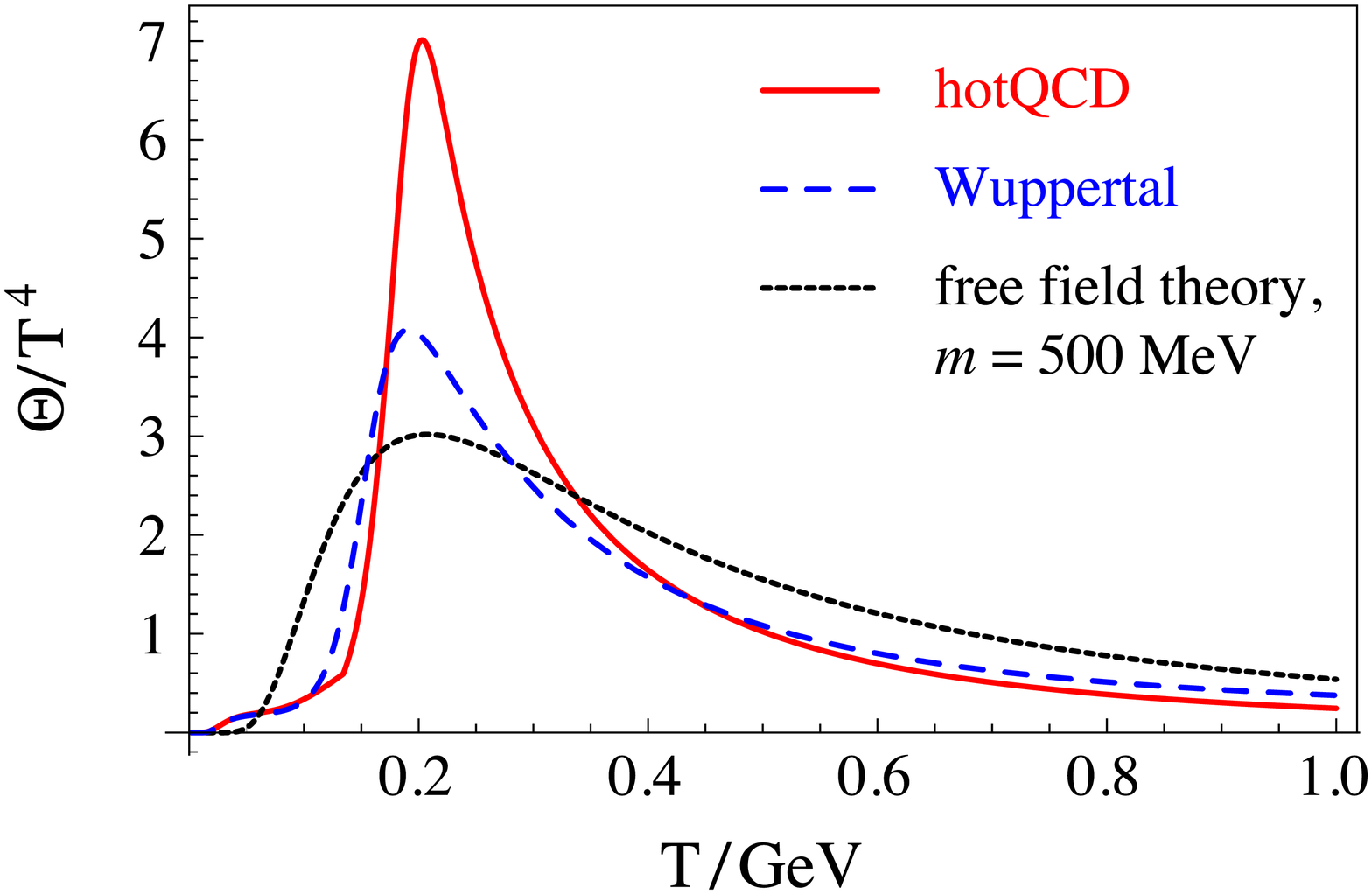}
\end{center}
\caption{Comparison of the thermodynamics of lattice QCD and of a theory of free quarks and gluons.  In all cases, three quark species ($u$, $d$, and $s$) have been included.  The curve labeled ``hotQCD'' is based on Ref.~\cite{Bazavov:2009zn}, while the curve labeled ``Wuppertal'' is from Ref.~\cite{Borsanyi:2010cj}; see the main text for details.  The free field theory curve was obtained by setting the masses of all the quarks and gluons artificially to $500\,{\rm MeV}$.  This counter-factual assumption is made in order to obtain the narrowest possible peak in $\Theta/T^4$ in the vicinity of the QCD transition, illustrating that interactions have a significant effect beyond introducing dynamically generated masses.}
\label{fig:QCDcomparison}
\end{figure}

The results from \cite{Bazavov:2009zn} and \cite{Borsanyi:2010cj} both support the conclusion that the peak in $\Theta/T^4$ is narrower and higher than one can get from free field expressions: see Fig.~\ref{fig:QCDcomparison}.  Perhaps this is not too surprising given that the QCD transition is almost a second-order phase transition, blunted only weakly into a cross-over, whereas free field expressions based on a finite particle spectrum are necessarily far from describing sharp phase transitions.

As is evident from Fig.~\ref{fig:QCDcomparison}, there is still some question about just how high and narrow the peak in $\Theta/T^4$ is.  The main difficulty is that there are a number of different lattice actions, all of which have in principle the same continuum limit, but which can in practice give significantly different results at finite lattice spacing.  Extrapolating to the continuum based on finite lattice studies can never be fully systematic.  The hotQCD results shown in Fig.~\ref{fig:QCDcomparison} are representative of p4 lattice data with lattice spacing $a$ chosen so that $N_\tau \equiv 1/(aT) = 8$ in the Euclidean time direction.  Using instead an asqtad action results in a peak roughly $15\%$ lower.  The Wuppertal results shown in Fig.~\ref{fig:QCDcomparison} are based on a stout-improved staggered fermion action with $N_\tau$ in the range of $8$ to $12$.  (For details on the p4, asqtad, and stout-improved staggered fermion actions, see \cite{Bazavov:2009zn,Borsanyi:2010cj} and references therein.)
Going forward, we will use the p4 results of \cite{Bazavov:2009zn}; not that we endorse them as more accurate, but instead because they are the most different from free field results, providing us with an upper limit of how sharp the QCD transition might plausibly be as a result of interactions.

In order to obtain a complete account of the visible sector degrees of freedom in the vicinity of the QCD transition, we combine the energy density and pressure obtained from (\ref{BazavovForm}) and (\ref{pFromTheta}) with free field treatments of all the leptons, and also the $c$ and $b$ quark, where all particles are constrained to have the same temperature.  A free field treatment is obviously not perfect (particularly for the $c$ quark), but improved approximations would be complicated.

%%%%%%%%%%%%%%%%%%%%%%%%%%%%%%%%%%%%%%%%%%%%%%%%%%%%%%%%%%%
\section{The electroweak phase transition}
\label{Electroweak}

\begin{figure}[h]
\begin{center}
\includegraphics[scale=0.7]{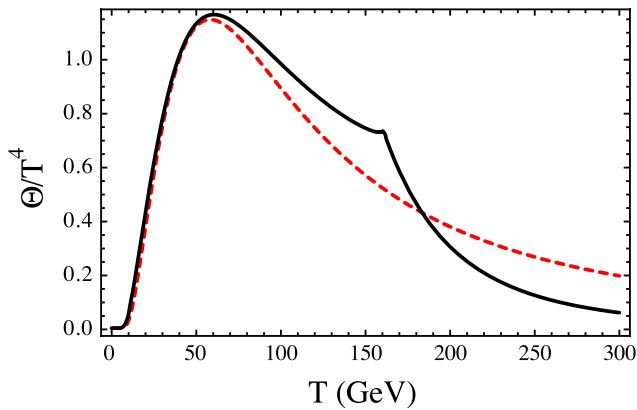}
\end{center}
\caption{Comparison of the ring-improved one-loop treatment of electroweak thermodynamics \cite{Arnold:1992rz} with free field expectations, shown by the solid black and dashed red lines, respectively.  Both curves are based on the minimal Standard Model with the Higgs of mass $m_H=125\,{\rm GeV}$  \cite{ATLAS:2012gk,CMS:2012gu}. The sharp feature at $160\,{\rm GeV}$, corresponding to the temperature at which the symmetry of the Higgs potential is broken, is a shortcoming of the ring-improved one-loop treatment.}\label{fig:ElectroweakComparison}
\end{figure}

It is far from obvious that the electroweak transition can be approximated by free field thermodynamics.  The masses of all observed Standard Model particles owe their existence to a non-zero Higgs expectation value, $\phi = \sigma \equiv 246\,{\rm GeV}$.  But this expectation value is eventually driven to zero at high temperatures.  An electroweak scale contribution to the cosmological constant accompanies this change in the Higgs expectation value.  Meanwhile, electroweak interactions introduce thermal corrections to particle masses.  Without accounting properly for these thermal corrections and other loop effects, the Higgs mass itself, now known to be approximately $125\,{\rm GeV}$ in vacuum 
\cite{ATLAS:2012gk,CMS:2012gu}, would become imaginary once the Higgs expectation value falls below the point where the tree-level potential is concave up.  One of the main conclusions of this section, illustrated in Fig.~\ref{fig:ElectroweakComparison}, is that, close to the peak of $\Theta/T^4$,
free field thermodynamics based on the vacuum particle spectrum nevertheless provides a decent approximation to the ring-improved one-loop treatment of \cite{Arnold:1992rz}, which is the simplest account of the electroweak transition that avoids obvious inconsistencies such as imaginary masses.  At substantially higher temperatures, we will show that the ring-improved one-loop treatment predicts a negative value of $\Theta/T^4$.

A more modern understanding of the electroweak transition \cite{Kajantie:1996mn}, based in part on lattice simulations, is that the transition is not weakly first order, as predicted by the ring-improved one-loop treatment, but is instead a cross-over. If anything, we expect the full non-perturbative results for the trace $\Theta$ of the stress tensor to be closer to the free field results than the ring-improved one-loop results are, though it is likely that the ring-improved one-loop treatment is still a good guide well above and well below the cross-over.

The treatment of \cite{Arnold:1992rz} proceeds in three steps:
 \begin{enumerate}
  \item First one produces thermally improved formulas for all the fields using self-energy diagrams.  The schematic form of these masses is $m^2(\phi,T) = m^2_{\rm tree}(\phi) + g^2 T^2$, where $g$ is a gauge coupling and $\phi$ is the Higgs field expectation value.  The precise forms of all the masses are listed in table~\ref{Table:ThermalMasses}.
  \item Next one assembles an effective potential, correct through one-loop order, as follows:
  \begin{widetext}
 \begin{eqnarray}
  V(\phi,T) &=& V_{\rm tree}(\phi) + V^{(1)}_{\rm vac}(\phi,T) + 
      V_{\rm therm}(\phi,T)  \\
  V_{\rm tree}(\phi) &=& {\lambda_0 \over 4} (\phi^2-\sigma^2)^2  \nonumber \\
  V^{(1)}_{\rm vac}(\phi,T) &=& \sum_j {s_j \over 64\pi^2} \Big[
   m_j(\phi,T)^4 \log {m_j(\phi,T)^2 \over m_j(\sigma,T)^2}   - {3 \over 2} m_j(\phi,T)^4 + 2 m_j(\phi,T)^2 m_j(\sigma,T)^2   - {1 \over 2} m_j(\sigma,T)^4 \Big]  \nonumber \\
  V_{\rm therm}(\phi,T) &=& \sum_j {g_j s_j \over 2\pi^2} T^4
    \int_0^\infty dx \, x^2  \log(1 - s_j e^{-\sqrt{x^2+m_j(\phi,T)^2/T^2}}) \,, \nonumber
 \end{eqnarray}
 \end{widetext}
where the sums over $j$ run over the entries in table~\ref{Table:ThermalMasses}, and $s_j=\pm 1$ for bosons and fermions, as in (\ref{FreeE}) and (\ref{FreeP}).  $V^{(1)}_{\rm vac}$ is the one-loop vacuum correction to the effective potential, and its dependence on $T$ arises only because, by prescription, one replaces the vacuum masses by the thermally improved mass formulas.  $V_{\rm therm}(\phi,T)$ is the one-loop contribution from thermal occupation numbers for all the fields.
  \item At any fixed temperature $T$, let $\phi(T)$ be the value of the Higgs field which minimizes $V(\phi,T)$.  Then the pressure is $p = -V(\phi(T),T)$.  The energy density can be extracted from the first law formula $\rho = -p + T \, dp/dT$.
 \end{enumerate}

\begin{table}
 \begin{ruledtabular}
 \begin{tabular}{c|c|c}
  \lower5pt\hbox{Particle } & Mass squared & Multiplicity \\[-2pt] 
   & ($m^2$) & ($g$) \\[4pt] \hline & & \\[-7pt]
  $W$ & $m_{W,\rm vac}^2 {\phi^2 \over \sigma^2}$ & $4$  \\[10pt]
  $W_L$ & $m_{W,\rm vac}^2 \left( {\phi^2 \over \sigma^2} + 
    {22 \over 3} {T^2 \over \sigma^2} \right)$ & $2$  \\[10pt]
  $\gamma$ & $0$ & $1$  \\[10pt]
  $Z$ & $m_{Z,\rm vac}^2 {\phi^2 \over \sigma^2}$ & $2$  \\[10pt]
  \lower10pt\hbox{$\gamma_L$} & ${1 \over 2} (g_1^2+g_2^2) C(T)$\hskip1.2in\ & 
    \lower10pt\hbox{$1$}  \\[-5pt]
        & $\quad{} - {1 \over 4} \sqrt{g_1^2 g_2^2 \phi^4 + 4 (g_1^2-g_2^2)^2 C(T)^2}$ & 
          \\[10pt]  
  \lower10pt\hbox{$Z_L$} & ${1 \over 2} (g_1^2+g_2^2) C(T)$\hskip1.2in\ & 
    \lower10pt\hbox{$1$}  \\[-5pt]
        & $\quad{} + {1 \over 4} \sqrt{g_1^2 g_2^2 \phi^4 + 4 (g_1^2-g_2^2)^2 C(T)^2}$ & 
          \\[10pt]
    & $C(T) \equiv {1 \over 4} \phi^2 + {11 \over 6} T^2$ &  \\[10pt]
  \lower10pt\hbox{$H$} & $\lambda_0 (3\phi^2 - \sigma^2) + {1 \over 2} \lambda_0 T^2$\hskip0.85in\  & 
     \lower10pt\hbox{$1$}  \\[-5pt]
        & $\ \ {} + 
           {1 \over 4} {T^2 \over \sigma^2} (2 m_{t,\rm vac}^2 + 2 m_{W,\rm vac}^2 + 
            m_{Z,\rm vac}^2)$ &  \\[10pt]
  $t$ & $m_{t,\rm vac}^2 {\phi^2 \over \sigma^2}$ & 12  \\[7pt]
  other & \lower5pt\hbox{$0$} & \lower5pt\hbox{$78$}  \\[-5pt]
  fermions & & \\[5pt]
  gluons & 0 & 16  \\[5pt]
 \end{tabular}
 \end{ruledtabular}
 \caption{Thermally corrected masses of Standard Model particles as a function of the Higgs expectation value $\phi$ and the temperature $T$, following \cite{Arnold:1992rz}.  A few entries require explanation.  At $T=0$, $\gamma_L$ becomes a physical, massless photon, but at $T \neq 0$ it mixes with the Higgs scalars along with the longitudinal $W_L$ and $Z_L$.  The $H$ entry corresponds to quanta of the magnitude $\phi$ of the Higgs doublet.  We ignore the masses of all the fermions except the top quark, and we also ignore strong interactions.\label{Table:ThermalMasses}}
\end{table}

\begin{table}
 \begin{ruledtabular}
 \begin{tabular}{c|c|c|c}
  \hskip0.1in\ Quantity\hskip0.2in\  & \hskip0.1in\ Value\hskip0.2in\  & \hskip0.1in\ Quantity\hskip0.2in\  & \hskip0.1in\ Value\hskip0.2in\  \\[2pt] \hline & & & \\[-10pt]
  $\sigma$ & $246\,{\rm GeV}$ & $\lambda_0$ & 
    ${m_H^2 \over 2\sigma^2}$  \\[2pt]
  $g_1$ & $0.357$ & $g_2$ & $0.652$  \\[2pt]
 \end{tabular}
 \end{ruledtabular}
 \caption{Standard Model parameters.\label{Table:Parameters}}
\end{table}
    
In Fig.~\ref{fig:ElectroweakComparison}, we compare the results of the ring-improved one-loop approach to the case of free field thermodynamics based on the masses and multiplicities of the Standard Model particle spectrum at zero temperature.  The kink in $\Theta/T^4$ at $160\,{\rm GeV}$ shows the location of the first order transition predicted by the ring-improved one-loop approach.  This kink is actually a very slight discontinuity in $\Theta/T^4$ as a function of $T$. The kink is smoothed out by the non-perturbative effects that make the entire transition a crossover.  
It is clear from Fig.~\ref{fig:ElectroweakComparison} that electroweak interactions and symmetry restoration have essentially no effect on the peak value of $\Theta/T^4$, and that they are mainly significant on the high-temperature side of this peak.  In the next two paragraphs we continue our investigation of the high-temperature behavior analytically.

At temperatures high above the electroweak transition, we expect that the leading contribution to the trace is due to the most massive relativistic species in the thermal bath. In the electroweak symmetric phase, however, the masses of most particles vanish. Although the longitudinal gauge bosons $\gamma_L,\,W_L,\,Z_L$ have large, thermally-improved masses, these temperature-dependent corrections drop out when we evaluate $\rho - 3 p = -4 p + T dp/dT$. This is because their thermally-improved masses, in the high-temperature, symmetry-restored phase, are dimensionless constants times the temperature, resulting in a contribution to $p$ that scales precisely as $T^4$.  The only particle in the spectrum whose mass in the symmetry-restored phase is {\it not} linear in the temperature is the Higgs.  So its thermal excitations dominate $\Theta$ at high temperatures.  An important sub-dominant contribution is the classical Higgs potential.  Altogether, at $T \gg \sigma$,
 \begin{equation}
  \Theta = h_2 T^2 + h_0 + {\cal O}(T^{-2}) \,, \label{ThetaExpand}
 \end{equation}
where
 \begin{eqnarray}
  h_0 &=&\frac{1}{2} \sigma^2 m^2_{H}  + \hbox{(one loop)}  \\
  h_2 &=& -\left( \frac{m_H}{2 \pi}\right)^2  \int_0^\infty dx \, \frac{x^2}{\sqrt{x^2+q^2} ({\rm e}^{\sqrt{x^2+q^2}}-1)}  \nonumber \\
  q^2 &=& ({m_H^2}+2{m^2_{t,\rm vac}}+2{m^2_{W,\rm vac}}+{m^2_{Z,\rm vac}})/(2\sigma)^2. \nonumber
 \end{eqnarray}
The contribution to $h_0$ shown explicitly is from the tree-level Higgs potential.  The corrections, indicated by $\hbox{(one loop)}$, are numerically small given the measured value of the Higgs mass.  The $h_2 T^2$ term is from the thermal excitations of the Higgs. Numerically, $-h_2 \sim m_H^2/45 \sim  (19\,{\rm GeV})^2$, valid for $T\gg \sigma$. In the absence of contributions beyond the Standard Model, the trace $\Theta$ is negative at high temperatures, starting at about the temperature $T_*  = \sqrt{-h_0/h_2}\sim 1200\,{\rm GeV}$ where the negative $h_2 T^2$ term in (\ref{ThetaExpand}) overtakes the positive $h_0$ term. The most negative value attained by $\Theta/T^4$ occurs at $T = \sqrt{-2 h_0/h_2} \simeq 1600\,{\rm GeV}$ at which point $g_*-g_{*,p} = -15 h_2^2/(2 \pi^2 h_0) \simeq -2\times 10^{-4}$. Because this negative trace is a result found using the ring-improved one-loop approximation, it is not to be regarded as iron-clad.  We leave for future work the interesting question of whether it persists beyond this approximation.

\begin{figure}[t]
\begin{center}
\includegraphics[scale=0.37]{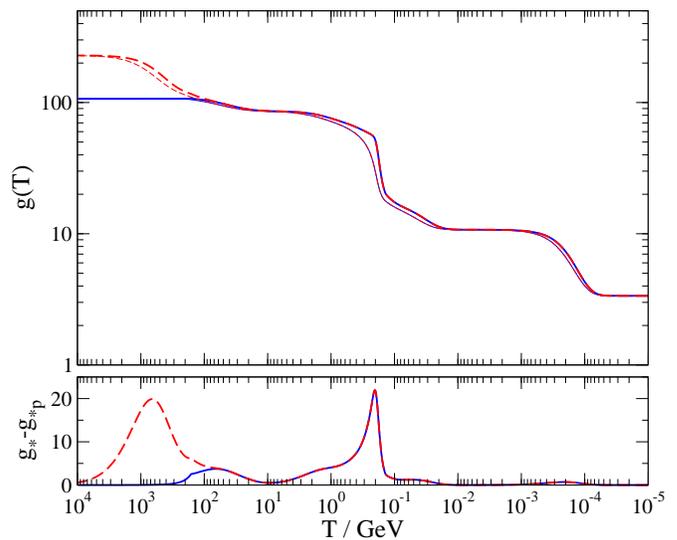}
\end{center}
\vspace{-0.5cm}
\caption{The degrees of freedom of the Standard Model and a supersymmetric extension are illustrated as functions of equilibrium temperature. In the top panel, the degrees of freedom of the energy density, $g_*$ (heavy curves), and pressure $g_{*p}$ (light curves), are shown for the Standard Model (solid curves) and a representative supersymmetric extension (dashed curves). In the lower panel, the difference $g_* - g_{*p}$ is shown. Note that adding a single dark matter species to the Standard Model curves does not produce a discernible change in these figures.}
\label{fig: DegreesOfFreedom}
\end{figure}

\begin{figure*}[t]
\begin{center}
\includegraphics[scale=0.66]{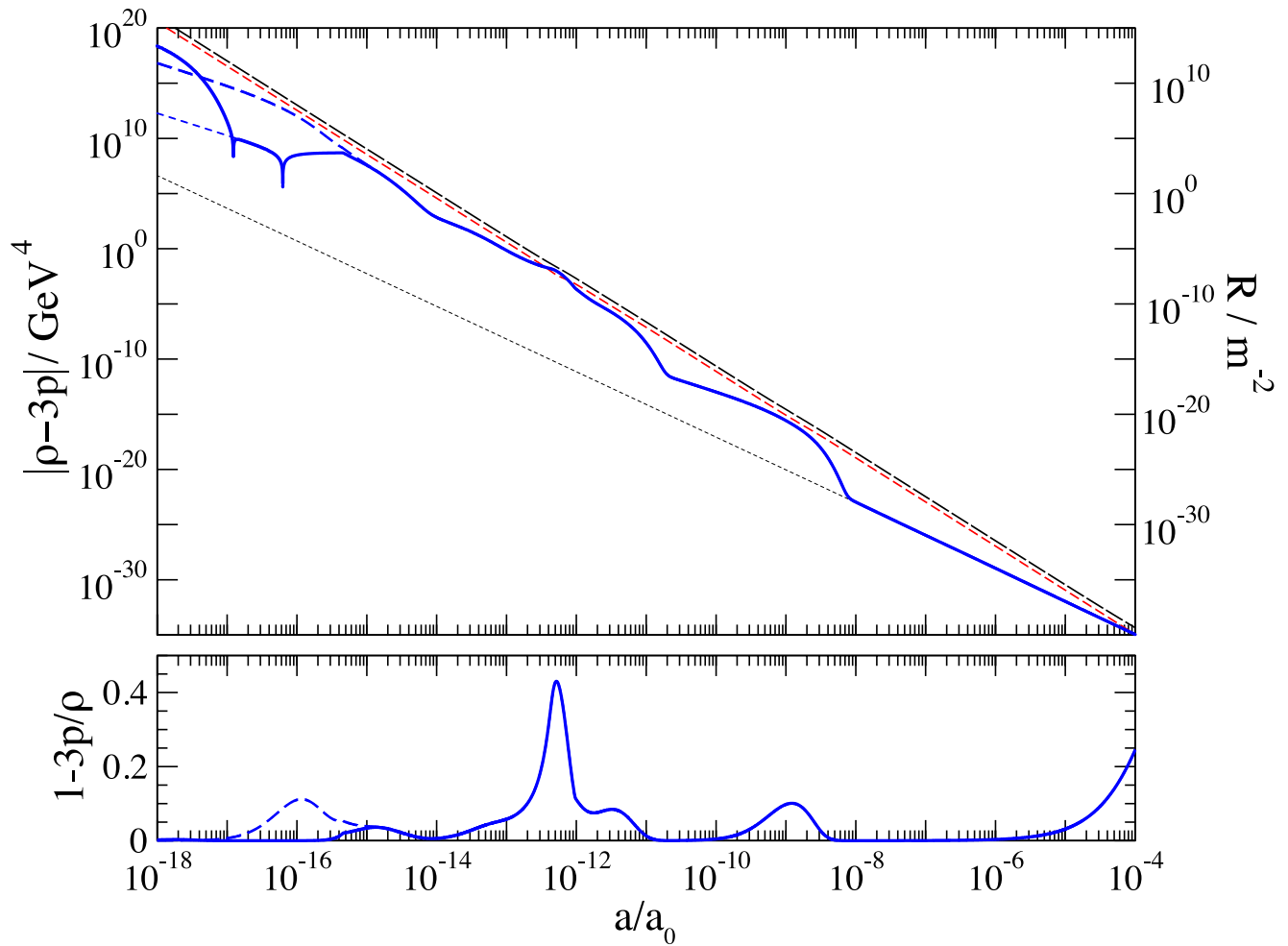}
\end{center}
\caption{
The Ricci scalar curvature in units of ${\rm m}^{-2}$, or equivalently the trace of the cosmological fluid stress-energy tensor in units of ${\rm GeV}^4$, as in Fig.~1, for $10^{-18} < a/a_0 < 10^{-4}$.  The energy density and the pressure are also shown as long dashed (black) and short dashed (red) lines.  Except as noted below, $\Theta = \rho-3p$ is positive.\\  For reference, a continuation of the energy density in pressureless matter is shown by the dotted black line; naive treatments that equate the trace with the matter density at early times are easily off by $\sim 5-10$ orders of magnitude. The solid (blue) curve shows $\Theta$ for the Standard Model plus a dark matter particle of mass $m_{DM} = 100\,{\rm TeV}$. The thin dashed (blue) curve shows the case of the SM through the electroweak transition as described in the text, with $\Theta$ becoming negative at $a/a_0 < 10^{-16}$. The long dashed (blue) curve near the left edge of the plot shows the case of the cMSSM. In the lower panel, the trace in units of the critical density is shown; the spikes at $a/a_0 = 10^{-16},\, 10^{-15},\, 4\times 10^{-13},\, 10^{-9}$  correspond to particles of the cMSSM becoming non-relativistic, the  electroweak transition, the QCD transition, and the departure of electrons from equilibrium as they become non-relativistic, respectively.}
\label{fig:TraceRadEra}
\end{figure*}

We may reasonably expect the LHC or direct dark matter detection experiments to discover new particle species at higher masses. Consequently, we supplement the Standard Model particle spectrum with a dark matter (DM) particle. Until it freezes out, a single, relativistic, bosonic species contributes $\Theta_{DM} = m_{DM}^2 T^2/12$ to the trace. If a bosonic dark matter particle has a mass $m_{DM} \lesssim \sqrt{12 h_2} \simeq 70\,{\rm GeV}$ or $m_{DM} \gtrsim 20\,{\rm TeV}$ then there will be a window in which the collective stress-energy trace is negative due to the Standard Model Higgs. Such a possibility is reflected in Fig.~\ref{fig:TraceRadEra}, where we include a $100\, {\rm TeV}$ dark matter particle which freezes out after it becomes non-relativistic, in our extrapolation of $\Theta$ back to times before and temperatures above the Standard Model electroweak transition.
  
In hindsight, it is perhaps not too surprising that a weakly coupled theory like the Standard Model manages to have thermodynamic properties not too far from free field theory.  We therefore feel justified to include here a similar free field account of the constrained Minimal Supersymmetric Standard Model (cMSSM), in which we use DarkSUSY \cite{Gondolo:2004sc} with input parameters corresponding to the best fit to the recent LHC data (Table 4 of Ref.~\cite{Strege:2012bt}: $m_0=389.51\,{\rm GeV}$, $m_{1/2}=853.03\,{\rm GeV}$, $A_0=-2664.79\,{\rm GeV}$, $\tan\beta=14.5$, and ${\rm sgn}(\mu)=+1$) to generate a zero-temperature mass spectrum, which we then plug into (\ref{FreeE}) and (\ref{FreeP}) without regard to the modifications that will arise from interactions and symmetry restoration.  Though these modifications are probably significant on the high-temperature side of the peak in $\Theta/T^4$, the main uncertainty lies not in their precise determination, but our ignorance of what the superpartner spectrum really looks like (assuming it is there at all).  As seen in Figs.~\ref{fig: DegreesOfFreedom} and \ref{fig:TraceRadEra}, an anomalous trace arises as the spectrum of sparticles with masses in the range $\sim 0.3 - 2$~TeV become non-relativistic.  More specifically: the best-fit model we used as a benchmark has a neutralino LSP at $363\,{\rm GeV}$, additional color singlet sparticles below $1\,{\rm TeV}$, and colored sparticles above $1.4\,{\rm TeV}$ except for a stop at approximately $817\,{\rm GeV}$.
  
Proceeding to ever higher energies requires a theory to guide our calculation of the trace. In the approach to GUT-scale energies, we may pass through a desert, or there may be new particle species littering the landscape. We should also expect couplings to run, although this is a small effect in simple extensions of the Standard Model.

Pausing to reflect on Fig.~\ref{fig:TraceRadEra}, we may assess the degree of error commonly made in the literature by assuming the curvature to be negligible, or else solely determined by the non-relativistic matter content extrapolated back from the matter-dominated era. Instead, we see that at various epochs, when the number of degrees of freedom changes sharply, the trace rises to an appreciable fraction of the total energy density.

%%%%%%%%%%%%%%%%%%%%%%%%%%%%%%%%%%%%%%%%%%%%%%%%%%%%%%%%%%%
\section{Inflation and reheating}

The history of curvature at earlier times requires a theory of the early Universe, which we take to be the inflationary dynamics of a scalar field. Whereas other theories of the primordial Universe will yield different predictions for the trace and curvature, the merit of inflation is the consistency with observations and ubiquity of examples. The generic course of inflation is exponential expansion proceeding with $p_i \simeq -\rho_i$ caused by the slow roll of the inflaton $\psi$ down its potential until some late stage when it breaks from slow roll and plummets down the potential towards the vacuum. In broad terms, the inflaton undergoes damped evolution as it decays into a turbulent mixture of relativistic particles and massless radiation. It is only after some transient period that these decay products equilibrate and can be appropriately described as a thermal bath with temperature $T$. The description of the cosmological fluid, trace and curvature, can then merge with our description of the high temperature limit beyond the Standard Model.

\begin{figure*}[t]
\begin{center}
\includegraphics[scale=0.66]{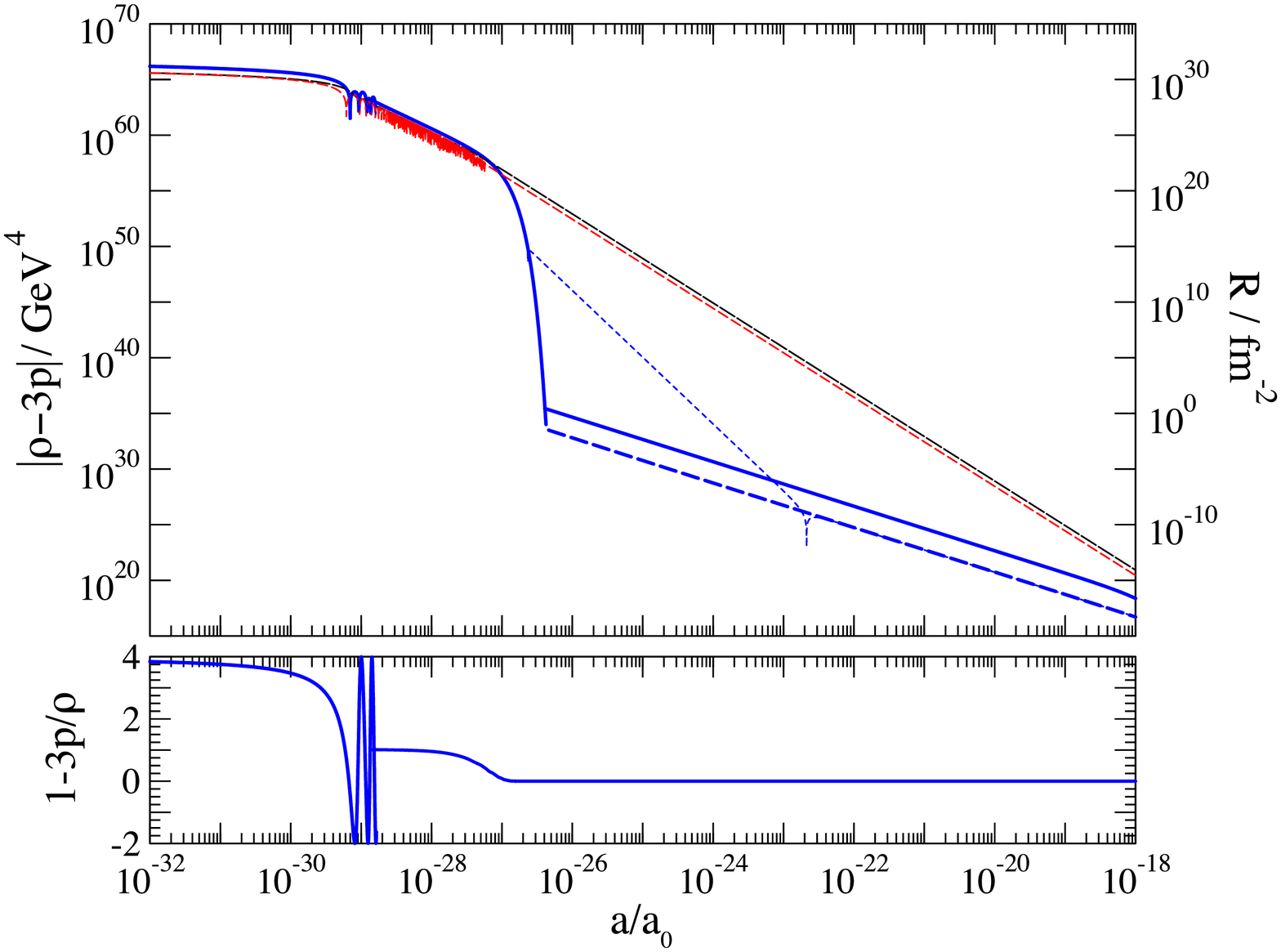}
\end{center}
\caption{The Ricci scalar curvature in units of ${\rm fm}^{-2}$, or equivalently the trace of the cosmological fluid stress-energy tensor in units of ${\rm GeV}^4$, as in Fig.~\ref{fig:TraceMattEra}, for $10^{-32} < a/a_0 < 10^{-18}$.  Except as noted below, $\Theta = \rho-3p$ is positive.  The energy density and the absolute value of the pressure are also shown as long dashed (black) and short dashed (red) lines.\\  Inflation continues at earlier times than shown in the plot, with all curves continuing as horizontal lines.  The solid (blue) line shows the extrapolation of the SM with a DM species with mass $100\,{\rm TeV}$ back to the post-inflationary epoch.  The long dashed (blue) line shows the extrapolation of the cMSSM where the LSP is at $363\,{\rm GeV}$. In the cMSSM scenario, a putative decay product of the inflaton having equation-of-state $+1$ is considered, for which the contribution to the trace is negative. In such a case, the curvature and trace may become negative, as shown in the range $3 \times 10^{-27} < a/a_0 < 3 \times 10^{-23}$ by the thin dotted (blue) line.  $\Theta$ also becomes negative during oscillations at the end of inflation.  In the lower panel, the trace in units of the critical density is shown. After the inflaton makes several oscillations across the bottom of its potential, and until the Universe is fully radiation dominated, we plot the time average of the trace for clarity of display.}
\label{fig:TraceInflEra}
\end{figure*}

A detailed analysis of the post-inflationary, preheating epoch including the approach to equilibrium requires a numerical simulation of the nonlinear, quantum fields ({\it e.g.} Refs.~\cite{Micha:2002ey,Berges:2004ce,Podolsky:2005bw}). Nonetheless, we can capture the basic behavior with a semi-analytic model. To be definite, we consider an inflaton with potential $V=\frac{1}{2}m^2 \psi^2$, which decays into relativistic particles and massless radiation via a damping term. Following Refs.~\cite{Kolb:1990vq,Turner:1983he}, we evolve the system 
\begin{eqnarray}
\ddot\psi + (3H+\Gamma)\dot\psi + V' &=&0, \\
\dot\rho_r + 4 H\rho_r - \Gamma\dot\psi^2 &=&0, \nonumber
\end{eqnarray} 
with $\rho_i = \frac{1}{2}\dot\psi^2 +V$, $p_i  = \frac{1}{2}\dot\psi^2 -V$. The field is started in a slow-roll state, with $m\sim 10^{14}$~GeV chosen to yield satisfactory inflation \cite{Kinney:2008wy,Mortonson:2010er,Martin:2010hh}. As slow roll ends, the inflaton begins to oscillate about the potential minimum, such that $\rho_i - 3p_i$ swings between $+4\rho_i$ and $-2\rho_i$;  despite the fact that the time-averaged inflaton pressure is small compared to the energy density, the trace and curvature become negative. This pressureless phase is shortlived, as an assumed inflaton decay width $\Gamma\sim10^{10}$~GeV causes the field to decay exponentially with cosmic time; the trace $\Theta_i = \rho_i - 3 p_i$ plummets until it meets the anomalous trace $\rho_r - 3 p_r \propto m^2 T^2$ due to any massive species among the inflaton decay products. This scenario, beginning from the end stages of inflation, is illustrated in Fig.~\ref{fig:TraceInflEra}.

The variety of inflationary models, and the challenges of accurately simulating the nonlinear dynamics of reheating, translates into a wide degree of uncertainty in the behavior of the trace and curvature in the approach to equilibration. In the above model we have assumed an idealized description of the radiation produced by the decay of the inflaton. If the inflaton decay products do not achieve the relativistic equation of state $w=1/3$ until after a brief equilibration period, then the trace may be much larger. If one of the decay products includes a kinetic-energy dominated scalar field with equation of state $w=+1$ then the trace contains a residual, negative term that decays as $\propto a^{-6}$.  This possibility is also illustrated in Fig.~\ref{fig:TraceInflEra}. 

The gap between the trace and the energy density is so wide in the interval between the end of inflation and the electroweak epoch, assuming no new physics, that we may question whether there are any other sources of curvature or trace. For example, long-wavelength fluctuations in the energy density and pressure might lead to an rms contribution. If the pressure and energy density perturbations imparted by inflation unto the cosmological fluid are characterized by a propagation speed $v$ where $v^2 = \delta p/\delta\rho$, then our question becomes: is $v^2$ as closely tuned to $1/3$ as is $w$, the background equation of state? In the case of perfectly adiabatic perturbations, with $v^2=w$, then the answer is yes. At some level, however, we expect a departure from adiabaticity. Recall that the linearized perturbation to the Ricci scalar in a coordinate system
\begin{equation}
ds^2 = -dt^2 +a^2(t) {\rm e}^{2 \zeta(t,\,\vec x)} \delta_{ij} dx^i dx^j
\end{equation}
is $\delta R = 2\delta R^{(3)} - 6(\ddot\zeta + 4 H \dot\zeta)$, and $\delta R^{(3)} =2\nabla^2\zeta/a^2$, where $\zeta$ is a familiar variable from studies of inflation. (See Ref.~\cite{Weinberg:2008zzc}.) For perturbations of wavelength much greater than the radiation-era Hubble radius, specifically those perturbations responsible for the observed large-scale structure, then $\zeta$ is a constant in time and $\delta R^{(3)}$ becomes negligibly small so that $\delta R \ll R$.  In this case, the homogeneous quantities we have calculated provide the leading contribution.

\begin{figure*}[h]
\begin{center}
\includegraphics[scale=0.66]{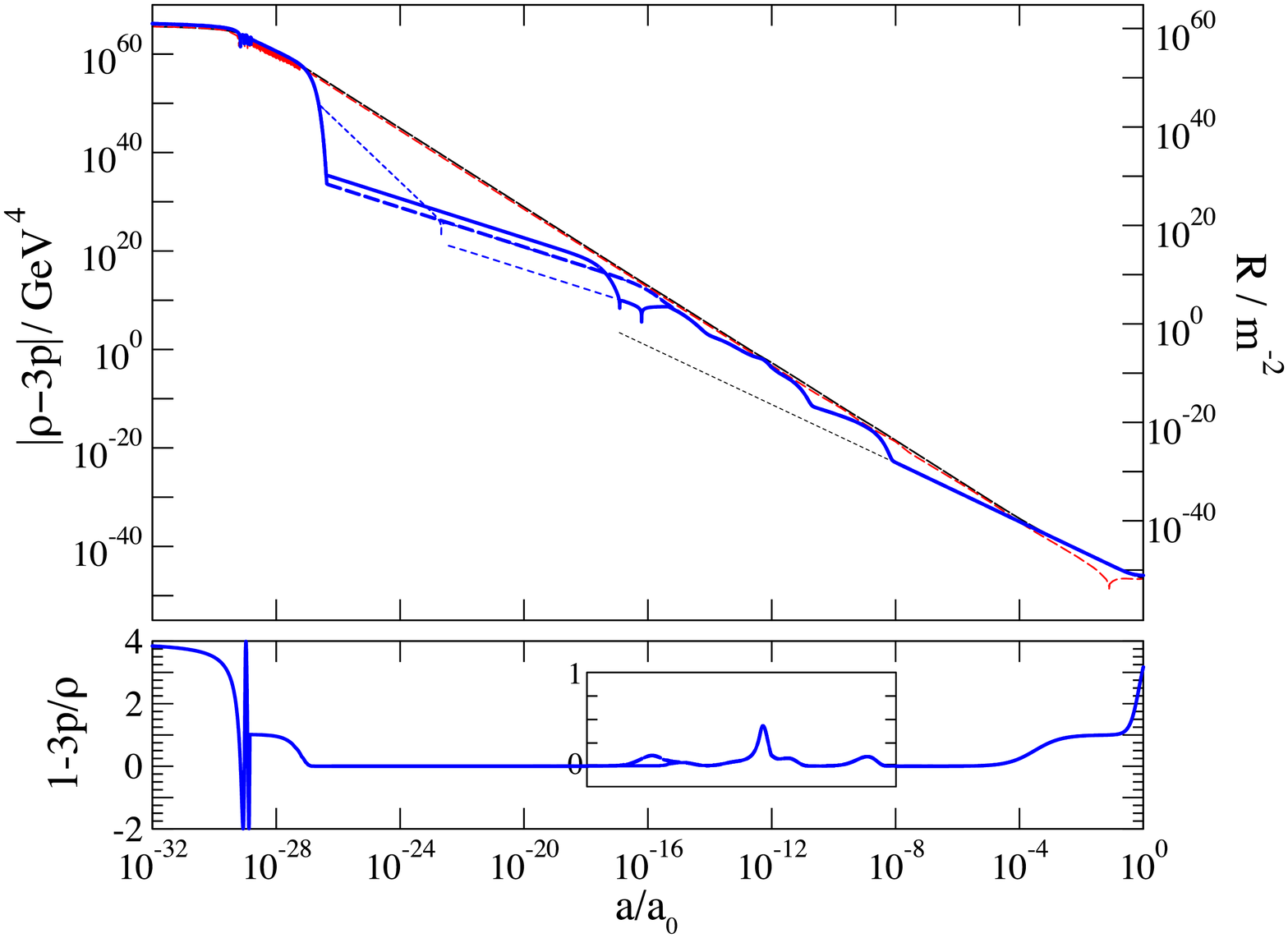}
\end{center}
\caption{A cosmic history of the Ricci scalar curvature in units of ${\rm m}^{-2}$, or equivalently the trace of the cosmological fluid stress-energy tensor in units of ${\rm GeV}^4$. The run of time scales is from the end stages of inflation to the present day, $10^{-32} < a/a_0 < 1$. The energy density and the absolute value of the pressure are shown by the thin long dashed (black) and thin short dashed (red) curves. Also shown is an extrapolation of the energy density in pressureless matter into the radiation era, for reference, which terminates near the possible freeze-out of a dark matter particle of mass $m_{DM} = 100\,{\rm TeV}$.  This plot is a composite of Figs.~\ref{fig:TraceMattEra}, \ref{fig:TraceRadEra}, and \ref{fig:TraceInflEra}, and the explanatory comments of previous captions apply to this figure as well.\\
In the lower panel, the trace in units of the critical density is shown.  The inset box is magnified by a factor of three to show spikes in curvature, due to the detailed thermal history of the SM and a supersymmetric extension, the electroweak transition, the QCD transition, and the departure from equilibrium of electrons. Simple treatments that attribute the curvature due to the contribution to the trace by the energy density in pressureless matter at early times underestimate the amplitude of the trace by many orders of magnitude, and miss out on episodes in which the curvature and trace are negative.}\label{fig:TraceFullHist}
\end{figure*}

To conclude this brief history of curvature, we show all three epochs in a single display, in Fig.~\ref{fig:TraceFullHist}.

%%%%%%%%%%%%%%%%%%%%%%%%%%%%%%%%%%%%%%%%%%%%%%%%%%%%%%%%%%%
\section{Discussion}

The history of the curvature and trace are of interest, beyond their intrinsic value as a record of the significant events in the large-scale evolution of our Universe. For example, an accounting of the thermal history of the universe is necessary to accurately account for the spectrum of inflation-produced gravitational waves \cite{Watanabe:2006qe}. However, the most dramatic effect may arise when considering new gravitational phenomena beyond general relativity.

Gravitational theories that introduce new degrees of freedom may require an accurate model of the trace of the stress-energy tensor. The canonical example is the nonminimally-coupled scalar field, for which the Lagrangian is 
\begin{equation}
{\cal L} =  \frac{1}{2\kappa^2}(1-2\kappa^2 \xi \varphi^2)R - \frac{1}{2}(\partial\varphi)^2 - V(\varphi) + {\cal L}_{SM},
\end{equation} 
and where ${\cal L}_{SM}$ includes the Standard Model. The $\varphi$ equation of motion is $\Box\varphi =  V' +  2\xi R \varphi$ so the curvature introduces a novel, time- and space-dependent effective mass such that when $\xi R < 0$ then the field is susceptible to a tachyonic instability ({\it e.g.}  \cite{Lima:2010xw,Lima:2010na}). The curvature is set by the dynamics of the field, which are in turn sourced by the trace of the stress-energy of any other forms of matter. Specifically,
\begin{equation}
R = \kappa^2 \frac{\Theta - (\partial\varphi)^2(12\xi-1)+4V-12 \xi \varphi V'}{1 - 2 \kappa^2\xi \varphi^2(1-12\xi)}.
\end{equation}
In the case of a strongly coupled field with $|\xi| \gg 1$, for which $|\xi \kappa \varphi| \ll 1$ so that the $\varphi$ field does not dominate the stress-energy, then we have approximately $R\simeq \kappa \Theta$. In the case of a conformally-coupled field, $\xi=1/12$ and $V=\lambda \varphi^4$, then $R=\kappa \Theta$, which means that the trace calculated in this paper is directly applicable to the evolution of $\varphi$: every feature in our $\Theta$ curve results in a feature in $\varphi$. Similarly, nonlocal quantum gravitational effects that depend upon quantities such as $\Box^{-1}R$ or $\Delta_p^{-1}R^2$, where $\Delta_p$ is a fourth-order conformally-invariant operator \cite{Mottola:2006ew,Deser:2007jk}, are sensitive to the history of curvature. Consequently, the results presented in this paper for the evolution of the trace may prove useful in detailed studies of new gravitational physics, such as gravitationally-coupled inflation \cite{Bezrukov:2007ep} or dark-energy motivated modifications of gravity ({\it e.g.} Ref.~\cite{Carroll:2004de}, or \cite{Caldwell:2009ix} and references therein).

%%%%%%%%%%%%%%%%%%%%%%%%%%%%%%%%%%%%%%%%%%%%%%%%%%%%%%%%%%%
\acknowledgments
We thank P.~Langacker, M.~Lisanti, and R.~Mawhinney for useful discussions.  The work of R.R.C.\ was supported in part by NSF PHY-1068027.  The work of S.S.G.\ was supported in part by the Department of Energy under Grant No.~DE-FG02-91ER40671.

\vfill
\clearpage

%%%%%%%%%%%%%%%%%%%%%%%%%%%%%%%%%%%%%%%%%%%%%%%%%%%%%%%%%%%

\end{document}